\documentstyle[12pt,amssym,aasms4,epsf]{article}
\received{June 18, 1999}
\accepted{December 8, 1999}
\slugcomment{Accepted for Publication in the Astrophysical Journal}

\lefthead{Kulkarni et al.}
\righthead{NICMOS Imaging of a Damped Ly-$\alpha$ Absorber at $z=1.89$}

\begin{document}

\title{NICMOS Imaging of the Damped Ly-$\alpha$ Absorber at $z=1.89$ \\
toward LBQS 1210+1731 : \\ Constraints on Size and Star 
Formation Rate}

\author{Varsha P. Kulkarni\altaffilmark{1}, John M. Hill, Glenn Schneider}
\affil{University of Arizona, Steward Observatory, Tucson, AZ 85721}
\author{ Ray J. Weymann, Lisa J. Storrie-Lombardi\altaffilmark{2},}
 \affil{Carnegie Observatories, Pasadena, CA 91101}
\author{Marcia J. Rieke, Rodger I. Thompson,} 
\affil{University of Arizona, Steward Observatory, Tucson, AZ 85721}
\and 
\author{Buell T. Jannuzi}
\affil{National Optical Astronomy Observatories, P. O. Box 26732, 
Tucson, AZ 85726-6732 }
\vskip -1.0in
\altaffiltext{1} {Present address: Clemson University, Dept. of Physics 
\& Astronomy, Clemson, SC 29634}
\altaffiltext{2} {Present address: SIRTF Science Center, Caltech, Pasadena, CA 91125}
\vskip 1.6in

\begin{abstract}
We report results of a high-resolution imaging search for the galaxy 
associated with the damped Lyman-$\alpha$ (DLA) absorber at $z=1.892$ 
toward the $z_{em}=2.543$ quasar LBQS 1210+1731, using HST/NICMOS.  
The images were obtained in the broad filter F160W and the 
narrow filter F190N with camera 2 on NICMOS, and were aimed at detecting 
the absorber in the rest-frame optical continuum and in H-$\alpha$ line 
emission from the DLA absorber. After suitable point 
spread function (PSF) subtractions, a feature is seen in both 
the broad-band and narrow-band images, at a projected separation 
of 0.25$\arcsec$ from the quasar. This feature may be associated with the 
DLA absorber, although we cannot completely rule out that it could be a 
PSF artifact. If associated with the DLA, the object would be 
$\approx 2-3$ $h_{70}^{-1}$ kpc in size with a flux of $9.8 \pm 2.4$ 
$\mu$Jy in the F160W filter, implying a luminosity at 
$\lambda_{central}=5500$ {\AA} in the rest frame of 
$1.5 \times 10^{10}$ $h_{70}^{-2}$ L$_{\odot}$ at $z=1.89$, 
for $q_{0}=0.5$. However, a comparison of the fluxes in the broad 
and narrow filters indicates that most of the flux in the 
narrow-band filter is continuum emission, rather than red-shifted 
H-$\alpha$ line emission. This suggests that if this object is the 
absorber, then either it has a low star formation rate (SFR), with 
a 3 $\sigma$ upper limit of 4.0 $h_{70}^{-2}$ M$_{\odot}$ yr$^{-1}$, 
or dust obscuration is important. It is possible that the 
H-$\alpha$ emission may be extinguished by dust, but this seems 
unlikely, given the typically low dust-to-gas ratios observed 
in DLAs. Alternatively, the object, if real, may be associated 
with the host galaxy of the quasar rather than with 
the damped Ly-$\alpha$ absorber. H-band images obtained with the 
NICMOS camera 2 coronagraph show a much fainter structure 
$\approx 4-5$ $h_{70}^{-1}$ kpc in size and containing four 
knots of continuum emission, located 0.7$\arcsec$ away from the quasar. 
This structure is not seen in images of comparison stars after similar 
PSF subtractions, and is also likely to be associated with the absorbing 
galaxy or its companions, although we do not know its redshift.  We have 
probed regions far closer to the quasar sight-line than in 
most previous studies of high-redshift intervening DLAs. The two 
objects we report mark the closest detected high-redshift DLA 
candidates yet to any quasar sight line. 
If the features in our images are associated with the DLA, they suggest 
faint, compact, somewhat clumpy objects rather than large, 
well-formed proto-galactic disks or spheroids. If the features are 
PSF artifacts, then the constraints on sizes and star-formation rates 
of the DLA are even more severe. The size, luminosity, and SFR estimates 
mentioned above should therefore be conservatively considered as 
upper limits. 

\end{abstract}

\keywords{quasars: absorption lines; galaxies: evolution; 
galaxies: intergalactic medium; infrared: galaxies; cosmology: 
observations}

%

\newpage
\section{INTRODUCTION}

Damped Lyman-$\alpha$ absorption systems detected in spectra of 
high-redshift quasars are believed to be the progenitors of present-day 
galaxies, because they show high H~I column densities (log $N_{HI} \ge 
20.0$) and display absorption 
lines of several heavy elements. However, there are various competing 
ideas regarding the nature of the galaxies underlying the DLAs. Wolfe 
et al. (1986) suggested that the DLAs are rotating proto-disks. This 
suggestion has also been made by Prochaska \& Wolfe (1997, 1998), based 
on asymmetric line profiles of the heavy-element absorption lines 
in DLAs. On the other hand, gas-rich dwarf galaxies have also been 
suggested as candidate objects for the DLAs (York et al. 1986; Matteucci, 
Molaro, \& Vladilo 1997). Recently, Jimenez, Bowen, \& Matteucci (1999) 
have suggested 
that high-redshift DLAs may arise in low-surface brightness galaxies. The 
lack of substantial chemical evolution found in studies of element 
abundances in DLAs (e.g., Pettini et al. 1999; Kulkarni, Bechtold, \& Ge 
2000a) also shows that the currently known population of DLAs seems 
to be dominated by metal-poor objects, so DLAs may consist of dwarf or 
low-surface brightness galaxies with modest star formation rates. 
Unfortunately, it is hard to determine what type of galaxies 
underlie the DLAs, since most previous efforts to 
directly image the high-redshift DLAs have failed. A few detections have 
been made at low redshifts, which showed those DLAs to arise in low 
surface-brightness galaxies (see, e.g., Steidel et al. 1995a, 1995b; 
LeBrun et al. 1997). But high-redshift DLAs with $z_{abs} < z_{em}$ 
have proven hard to detect, and the question of the 
nature of galaxies giving rise to these DLAs is still open. 

Many of the previous attempts to detect the emission from DLAs 
concentrated on the Ly-$\alpha$ emission, which is an expected signature 
from a star-forming region (e.g. Smith et al. 1989; Hunstead, Pettini, 
\& Fletcher 1990; Lowenthal et al. 1995). There have been only a few  
Ly-$\alpha$ detections of DLAs so far. M{\o}ller \& Warren (1998) and 
M{\o}ller, Warren, \& Fynbo (1998) detected Ly-$\alpha$ emission in the 
fields of two DLAs at $z=2.81$ and $z=1.93$. However, both of these 
DLAs have $z_{abs} \approx z_{em}$ and may be different from  
the general population of intervening DLAs. Djorgovski et al. (1996) 
and Djorgovski (1997) reported Ly-$\alpha$ emitting objects with 
$R \sim 25$ (and inferred SFRs of a few M$_{\odot}$ yr$^{-1}$) 
in fields of a few DLAs, located at 2-3 $\arcsec$ from the 
quasar. However, the Ly-$\alpha$ technique cannot definitively measure 
the star formation rates of the DLAs because of the generally 
unquantifiable effects of dust extinction in the systems. The lack of 
detections in the other 
Ly-$\alpha$ studies of interevening DLAs could indicate either that 
DLAs have low 
star formation rates (SFR) or that the emission is extinguished by dust. 
As pointed out by Charlot \& Fall (1991), even small quantities 
of dust are sufficient to extinguish the Ly-$\alpha$ emission, since 
resonant scattering greatly increases the path length of Ly-$\alpha$ 
photons attempting to escape from an H I cloud. Indeed, observations 
of reddening of background quasars and evidence for depletion of 
Cr, Fe, Ni etc. relative to Zn suggest 
the presence of a small amount of dust in DLAs (see, e.g., Pei, Fall, 
\& Bechtold 1991; Pettini et al. 1997; Kulkarni, Fall, \& Truran 1997). 
Thus, it is hard to constrain the SFRs in DLA galaxies using the 
non-detections or weak detections of Ly-$\alpha$ emission. 

The issues of dust and SFR in high redshift DLAs are also 
important in view of recent claims based on mid-IR and far-IR 
observations that a large fraction of the star formation at high 
redshifts is hidden from us by dust obscuration (e.g., Elbaz et al. 
1998; Clements et al. 1999). One way to discern whether the 
previous non-detections of Ly-$\alpha$ were due to low SFR or presence 
of dust is to search for longer wavelength emission lines less affected 
by dust extinction and not subject to resonant scattering.  The 
ground-based near-IR spectroscopic survey of Bunker et al. (1999), 
which searched for redshifted H-$\alpha$ emission in $11 \arcsec \times 
2.5 \arcsec$ regions around 6 quasars with DLAs at $z > 2$ and reached  
3 $\sigma$ detection levels of 6-18 M$_{\odot}$ yr$^{-1}$, failed to 
detect any redshifted H-$\alpha$ emission from the DLAs in their sample. 
Some of the ground-based narrow-band photometric surveys for H-$\alpha$ 
emission from DLAs have also failed to detect any emission line objects 
in the DLA fields (e.g., Teplitz, Malkan, \& McLean 1998, who however 
found H-$\alpha$ emitters in the fields of some weaker non-DLA metal 
line systems). Some other 
narrow-band searches for H-$\alpha$ emission have revealed multiple 
objects in the DLA fields separated by several arcseconds from the quasar 
(2-12$\arcsec$ for Bechtold et al. 1998, 9-120 $\arcsec$ for Mannucci 
et al. 1998). These surveys, which had 3 $\sigma$ detection limits 
of $\sim 5$  M$_{\odot}$ yr$^{-1}$ (Bechtold et al. 1998) or  
$\gtrsim 10$ M$_{\odot}$ yr$^{-1}$ (Mannucci et al. 1998), found the 
H-$\alpha$ emitting objects to have a wide range of inferred SFRs 
(10-20 M$_{\odot}$ yr$^{-1}$ for Bechtold et al. 1998, 
6-90 M$_{\odot}$ yr$^{-1}$ for Mannucci et al. 1998). The relatively 
large separations of these 
emission line objects from the quasars indicates that they are not 
the DLA absorbers themselves, but star-forming companions in the same 
larger structure (e.g. sheet or filament) as the DLA. None of these 
ground-based surveys has been able to probe the regions very close to 
the quasar sightline (angular separations $< 2$ $\arcsec$), because 
of the limitations imposed by seeing in these studies. While these 
studies offer interesting information about the environments of 
the DLAs, high sensitivity diffraction-limited imaging is necessary 
for the detection of the DLA absorbers themselves (to probe 
small angular separations), and thus for determining the morphology 
and SFRs of the DLAs. The HST WFPC2 study of Le Brun 
et al. (1997) has detected candidates with angular separations $< 2$  
$\arcsec$ in broad band images for six 
DLAs at $z < 1$ and one DLA at $z=1.78$. However, the information
obtained from this study about the nature of high-redshift DLAs is 
limited since no narrow-band images were obtained and since the sample 
contained only 1 DLA at $z > 1$. As mentioned earlier, the HST WFPC2 
study of M{\o}ller \& Warren (1998 and references therein) detected
Ly-$\alpha$ emission in a $z_{abs} > z_{em}$ DLA, but this DLA may
differ from intervening DLAs. 

To summarize, many previous attempts to detect emission from
high-redshift intervening DLAs have failed. The few detections so far 
consist mainly of either weak Ly-$\alpha$ detections (which cannot 
constrain the SFR completely) or detections of H-$\alpha$ companions 
at fairly large angular separations from the quasars. There are only 
four objects detected so far in fields of high-$z$ interevening DLAs at 
small angular separations. These objects have impact parameters between 
4.3 and 11.5 h$_{70}^{-1}$ kpc (where $H_{0} = 70$ h$_{70}$ km s$^{-1}$ Mpc$^{-1}$), and are promising candidates for the DLAs 
in those sightlines (see M{\o}ller \& Warren 1998 and references therein; 
the other DLA impact parameter data listed in M{\o}ller \& Warren 1998 
are biased toward $z_{abs} \approx z_{em}$ DLAs.). To further increase the number of promising candidates for high-redshift 
intervening DLAs, it is necessary to carry out more deep high spatial 
resolution near-infrared searches for DLAs. 

We have obtained deep diffraction-limited images of three DLAs at 
$z \sim 2$ with the Near Infrared Camera and Multi-Object Spectrometer 
(NICMOS) 
onboard the Hubble Space Telescope (HST). Here we describe our NICMOS 
observations of the quasar LBQS 1210+1731 ($z_{em}=2.543 \pm 0.005$; 
Hewett, Foltz, \& Chaffee 1995), which has a 
spectroscopically known damped Ly-$\alpha$ absorber ($z_{abs}=1.8920$ 
and log $N_{HI} = 20.6$; Wolfe et al. 1995). Our observations have the 
unique benefit of combining high near-IR sensitivity and high spatial 
resolution with a more stable and quantifiable PSF than is currently 
possible with ground-based observations. A further feature of some of 
our observations is the use of the NICMOS coronagraph, which greatly 
decreases the scattered light background outside of the coronagraphic 
hole and therefore allows a study of the environment of the DLA. Our 
analysis indicates two objects at 0.25 $\arcsec$ 
and 0.7 $\arcsec$ from the quasar that we cannot explain as any known 
artifacts of the PSF. We believe that these objects are likely to be 
real and may be associated with the DLA and its companions, at 
impact parameters of 1.5 and 3.8 h$_{70}^{-1}$ kpc. 
We have thus probed regions far closer to the quasar sight-line than 
in most previous studies of high-redshift intervening DLAs, and the two 
objects we report 
mark the closest detected high-redshift intervening DLA candidates yet 
to any quasar sight line. 
Sections 2, 3, and 4 describe the 
observations, data reduction, and the subtraction of the quasar point 
spread functions. Our results are described in section 5. 
Section 6 describes various tests of our data analysis procedures, 
carried out to investigate whether the features seen after PSF 
subtraction are real. A summary of the results of the various 
data analysis tests is given in subsection 6.12. (Readers interested 
mainly in the scientific discussion can go directly from section 5 
to subsection 6.12.) Finally, sections 7 and 8 discuss the implications  
of our observations for sizes, environment, and star-formation rates 
of DLA galaxies.

\section{OBSERVATIONS}

The field of LBQS 1210+1731 was first observed on 1998, July 22 from 
07:14 UT to 16:40 UT, using NICMOS camera 2 (pixel scale $\approx 0.076 
\arcsec$, field of view $19.45 \arcsec \times 19.27 \arcsec$). 
A sequence of spatially offset broad-band images was obtained in 
multiaccum mode with the F160W (H) filter (central wavelength 
1.5940 $\mu$m, FWHM 0.4030 $\mu$m). Field offsetting was accomplished 
with a 5-point spiral dither pattern in steps of $\approx 7.5$ pixels, 
using the Field Offset Mirror (FOM) internal to NICMOS. The exposures 
at each dwell point were 512 s long, giving a total integration time 
of 2560 s. See MacKenty et al. (1997) for a detailed description 
of NICMOS imaging modes and options. The multiaccum observations 
consisted of non-destructive readouts 
in the ``step32'' readout timing sequence, i.e. ``multiaccum'' readouts 
separated logarithmically up to 32 s and linearly in steps of 32 s beyond 
that. In addition, narrow-band images were obtained in the filter F190N 
(central wavelength 1.9005 $\mu$m, FWHM 0.0174 $\mu$m), 
in which the redshifted H-$\alpha$ emission from the DLA, if present,  
would lie.  Four-point spiral dither patterns in steps of 7.5 pixels, 
with a 704 s ``step64'' multiaccum exposure at each dwell point, were 
repeated in five successive orbits, resulting in a total integration 
time of 14,080 s.  The spatial resolution 
of the F160W and F190N images is 0.14 $\arcsec$ (1.8 pixels) and 0.17 
$\arcsec$ (2.1 pixels) FWHM, respectively. Thus, camera 2 is almost 
critically sampled at the wavelengths used for our observations. 

Finally, broad-band images in the F160W filter were also obtained 
using the camera 2 coronagraph on 1998, July 29 from 10:14 to 13:07 UT. 
These consisted of an initial pair of 92 s long target-acquisition 
images, which were followed by 
placement of the target in the coronagraphic hole (0.3 $\arcsec$  or 4 
pixels in geometrical radius) and then integration of the object for a 
total of 4960 s (5 exposures of 480 s each in the first orbit and 5 
exposures of 512 s each in the second orbit, all using the step32 
multiaccum timing sequence). No dithering was used, of course, for 
the coronagraphic observations. The NICMOS coronagraph is 
comprised of two optical elements, a 165 $\mu$m physical diameter hole 
in the camera 2 field divider mirror at the reimaged HST f/24 
optical telescope assembly (OTA) focus and a 
Lyot stop at a cold pupil in the cryostat. The coronagraphic system 
significantly reduces both scattered and diffracted energy from the 
occulted target's point spread function core by factors of 4-6, 
compared to direct imaging (Schneider et al. 1998; 
Lowrance et al. 1998). Thus our coronagraphic images have higher 
sensitivity than the non-coronagraphic images for detecting those   
foreground damped Ly-$\alpha$ absorber or associated companions  
that are much fainter than the quasar and lie outside the coronagraphic 
hole.  

To circumvent image artifacts known as ``bars'' in all our camera 2 
images, cameras 1 and 3 were run in parallel, as discussed by Storrs 
(1997).

\section{REDUCTION OF IMAGES}

The images were reduced using the IRAF package Nicred 1.8, developed 
specifically for the reduction of multiaccum NICMOS data (McLeod 
1997). The dark image used was that made from 
on-orbit dark exposures taken during the NICMOS calibration program. 
For the non-coronagraphic images, the flat-field image used was 
made from on-orbit exposures taken with the internal calibration 
lamps during the NICMOS Cycle-7 calibration program. For the coronagraphic 
images, the flat-field image was made with target-acquisition data 
taken just before the coronagraphic exposures. This ensures that 
the coronagraphic hole is in the same position on the detector for 
the flat as for the 
quasar data, which is critical for studying faint objects close 
to the edge of the coronagraphic hole. (The standard calibration 
flats are not adequate for this purpose because the position of the 
coronagraphic hole on the detector changed with time, and a 
flat exposure taken at another time had the hole in a different 
place.) 

First, the exposures at each individual dither position were reduced 
using Nicred 1.8. Briefly, the steps followed by Nicred 1.8 are as 
described below: 

1. subtraction of the zeroth read from successive reads, both for 
the quasar data and the dark data, 

2. dark subtraction, read by read,

3. linearity correction, cosmic ray rejection, and fitting of slope 
to the successive reads in the multiaccum data, to get count rates 
in ADU s$^{-1}$,

4. correction of non-uniform bias level across the array (``the 
pedestal effect''), 

5. repeating step 3 on the bias-corrected image to get more accurate 
count rates, 

6. flatfielding using the appropriate flats,

7. subtracting the median of each row from that row and then 
likewise for columns, to remove bands caused by bias jumps during 
simultaneous use of amplifiers of other cameras in parallel, and 
thus to improve the flatness of the background,

8. fixing bad pixels using bi-cubic spline interpolation across the 
neighboring pixels. 

The images for the different dither positions 
were registered by cross-correlating with the IRAF task xregister. 
The quasar was used as the reference object since no other point sources 
were available in our images. 
Finally, the registered images were averaged together using a bad-pixel 
mask that took out any remaining bad pixels, and rejecting pixels 
deviating by more than 3 $\sigma$ from the average of the five F160W 
images, using averaged sigma-clipping.

For the F190N images, where there were five exposures (one in each orbit) 
at each of the four dither positions, we first median-combined the five 
exposures at each position separately, and then registered and 
median-combined the four positions together to make the final image.
For the coronagraphic F160W images, where there were five exposures at 
the same position in each of the two orbits, we averaged the 
exposures in each orbit separately and then took a weighted average 
(weighting by exposure times) of the combined exposures from the 
two orbits. 

In an attempt to gain the higher spatial resolution made possible by 
the half-integral dithers (in steps of 7.5 pixels), we also experimented 
with magnification (repixelization) of the images at the 
individual dither positions before combining them. The images 
for each individual dither position processed as per steps 1-8 above 
were magnified (i.e. numerically resampled) by factors of two each in 
x and y directions. A cubic spline interpolation was used to divide the 
pixels into subpixels, with the flux kept conserved. As discussed 
further in section 6.7, our results do not depend much on whether or 
not the magnification is done.

Figs. 1, 2, and 3 show the final reduced images for the 
non-coronagraphic F160W, non-coronagraphic F190N, and coronagraphic 
F160W observations. The orientations of Figs. 1 and 2 agree 
exactly while they differ from that of Fig. 3 by only 2.026 degrees. 
All three images have an essentially zero background. The F190N 
image shows a weak residual flat field and nonuniformities in the 
corners caused by amplifier glow. This effect is much less noticeable 
in the reduced F160W image. We believe that the F190N image is limited 
by the quality of the F190N flat field available to us. 
The F190N flat field, made from six 192 s long in-flight exposures to 
calibration lamps, has a count rate of 37.72 ADU s$^{-1}$, while the 
F160W flat, made from nine 24 s exposures, has a count rate of 1113 
ADU s$^{-1}$. The rms deviation in the count rate per pixel is 
about $2 \%$ for each of the six frames combined to make the F190N flat, 
while it is about $0.02 \%$ for each of the nine frames combined to make 
the F160W flat. The lack of a better F190N flat is unfortunate. 
However, this should not be a serious problem for the quasar and DLA 
images, since they lie in the central part of the array. Figures 1 
and 2  show the quasar point source along with the diffraction pattern. 
The coronagraphic image in Fig. 3 shows the quasar light to be reduced 
greatly, although not completely. To study whether there is any 
additional underlying faint emission from the DLA in any of these 
images, we need to subtract the respective PSFs. 

\section{SUBTRACTION OF THE QUASAR POINT SPREAD FUNCTION}

\subsection{SELECTION OF THE PSF STAR}

Reference point spread functions for subtraction were obtained by using 
observations of stars in the same filter / aperture combinations as those 
employed for the quasar imaging. PSF star observations were not included 
in our own observations since we wanted to maximize the use of the 
available HST observing time for imaging of the quasar fields. We 
therefore used PSF star observations from other programs (in particular 
the stellar images from the photometric monitoring program carried out 
during Cycle 7 NICMOS calibration) for constructing the reference 
PSFs for subtraction. Such directly observed PSFs, when exposed to 
high S/N, are expected to provide better match to the quasar data than 
the theoretical Tiny Tim PSFs (since the observed PSFs incorporate any 
real optical effects not simulated 
in Tiny Tim). We have also actually experimented with the use of 
calculated Tiny Tim PSFs and find that they do indeed provide poorer 
match to the quasar than the observed stellar PSFs. 
 
For the non-coronagraphic images, the PSF observations 
were chosen such that the telescope focus ``breathing'' 
(Bely 1993) values matched as closely as possible the 
values for the DLA observations. This is 
important because changes in the HST focus translate into 
corresponding changes in the fine structure of the PSF. To estimate 
the OTA focus positions for the epoch of the quasar 
and PSF star observations, we used the HST focus ephemerides provided 
by STScI (Hershey 1998; Hershey \& Mitchell 
1998). For the non-coronagraphic F160W and 
F190N images, we used the PSF star P330E, observed on July 8, 1998 
and May 29, 1998 respectively. We also studied the effect of PSF 
variations on our results by using PSF observations of P330E with 
a range of different ``breathing'' focus positions 
obtained on different dates, and also by using observations 
of other PSF stars. (See section 6.2 and 6.3 for a detailed 
description.) The F160W non-coronagraphic image of P330E, made by 
combining four exposures of 3 s each, had a count rate of 108.10 
ADU s$^{-1}$ at the maximum of the first Airy ring. The corresponding 
quasar image, made by combining five exposures of 512 s each, 
had a count rate of 1.91 ADU s$^{-1}$ at the maximum of the first 
Airy ring. For the F190N filter, the P330E image, made by combining 
three exposures of 64 s each, had 0.055 ADU s$^{-1}$ at the maximum of 
the first Airy ring. The corresponding count rate was 0.0013 
ADU s$^{-1}$ for the F190N quasar image, made by combining 
20 exposures of 704 s each.

For the coronagraphic observations, the choice of the PSF 
star was guided by the requirement that the position of the 
star in the coronagraphic hole be as close as possible to 
that of the quasar in our observations. This is critical, because 
even when the target-acquisition flight software succeeds in acquiring 
the target and putting it in the coronagraphic hole, there are usually 
some small residual differences between the actual position where the 
target is placed and the desired position of the target in the hole, 
i.e. the ``low scatter point'' of the coronagraph (see 
Schneider 1998 for details). The PSF wings and ``glints'' 
from the edge of the coronagraphic hole depend sensitively 
on the precise position of the point source within the hole. 
We therefore used the observations of star GL83.1 for which 
we had coronagraphic observations (from another NICMOS GTO program), 
with the star placed at a position within 0.04 
pixels of the position of the quasar LBQS 1210+1731 in our 
data. The observations of GL83.1 were taken on 
August 1, 1998 at a breathing value close to that for our 
quasar coronagraphic observations. 

For all the ``primary'' PSF star choices, the proximity of 
the observation dates with those of our quasar observations 
also ensures that the plate scale of the camera is the same 
for the PSF and the quasar observations. 

\subsection{SUBTRACTION OF THE PSF STAR}

All of the observations of the PSF star P330E were obtained in 
4-point spiral dither patterns in steps of 4.0 $\arcsec$ 
($\approx 52.6$ pixels). The dithers for the PSF star were 
obtained by using actual spacecraft movements, while the dithers 
for LBQS 1210+1731 were obtained by moving the field 
offset mirror (FOM) internal to NICMOS. But the use of 
the FOM should not cause any differences between the combined 
quasar PSF and the reference star PSF, since we 
registered all of the quasar exposures individually to 
a common reference before combining. 
The PSF star observations were analyzed 
in exactly the same manner as the quasar observations, following 
the procedure outlined in section 3. The same interpolation scheme 
was used for resampling of the PSF star and quasar images. The 
difference in the dithering steps for the quasar and the PSF star 
may give rise to difference in actual sampling of the quasar and 
PSF star images. But, as 
described in section 6.7, we have verified that the 
difference images are reproduced well when both the 
quasar and PSF star images are numerically resampled by 
a factor of two. The final reduced PSF star 
images were subtracted from the corresponding 
quasar images after suitable scaling and registration,  
using the IDL program ``IDP-3'' (Lytle et al. 1999). 
The scale factors were chosen using the relative 
intensities of the PSF wings in the quasar image and 
the PSF star image. For the coronagraphic image, the 
relative intensities of the PSF ``glints'' near the 
edge of the hole were also used in determining the 
PSF scaling factor. All the parameters (i.e., relative 
x and y alignment of the PSF star image with respect 
to the quasar image and the intensity scaling factor 
for the PSF star image) were fine-tuned iteratively 
to obtain the minimum variance in roughly 
3 $\arcsec$  x 3 $\arcsec$ subregions (around the quasar) 
in the PSF-subtracted image. Radial plots of the quasar image, 
the aligned and scaled PSF image, and the difference of the two 
were also examined to check the alignment and scaling of the 
PSF. Figs. 4a, 5a, 6a show zoomed  
$\approx 3 $\arcsec$  \times 3 $\arcsec$ $ subregions around 
the quasar, from the non-coronagraphic F160W, non-coronagraphic 
F190N, and coronagraphic F160W images shown in Figs. 1, 2, 3, 
respectively. Figs. 4b, 5b, 6b show the PSF-subtracted versions 
of Figs. 4a, 5a, and 6a, respectively, using the closest 
matching PSFs available. 

\section{RESULTS}

\subsection{NON-CORONAGRAPHIC F160W IMAGES}

Fig. 4b shows the F160W image after subtraction of the PSF image of 
star P330E obtained on July 8, 1998. The 
fidelity of the PSF subtraction is seen from the fact that the 
diffraction pattern disappears completely and most of 
the residual image contains a random mixture of positive and negative 
values. The radially symmetric residuals may be explained, for the 
most part, by a mismatch  between the SED of the quasar and that of 
the PSF star (spectral type G2V). 
These color terms lead to small differences in the structure and 
size-scale of the PSFs. These differences are non-negligible under 
the $\approx 25 \%$ bandpass of the F160W filter, but are negligible 
under the 1$\%$ bandpass of the F190N filter. See section 6.3 for 
further discussion.  The main asymmetric residual 
is the emission feature to the ``lower right'' of the center, 
about 3 pixels (0.26 $\arcsec$) away from the center. (This feature 
is seen more clearly if the data are sub-sampled by a factor of 2, 
as discussed further in section 6.7 and Fig. 15.)
There is no correspondingly strong and symmetrically located negative 
feature in the image, and the bright knot can not be made to 
disappear after reregistration of the PSF and quasar images or 
rescaling of the PSF image without causing large negative residuals 
elsewhere (see section 6.9 and Fig. 18). 
We cannot completely rule out that this ``knot'' is an 
artifact in the PSF. However, given the significant excess over a 
number of pixels, it is likely that it is a real feature. 
This feature (which we name object ``O1'') is about 0.40 $\arcsec$ long. 
If this emission knot is associated with the damped Ly-$\alpha$ absorber 
at $z=1.892$, then it is $\approx$ 2.4 h$_{70}^{-1}$ kpc long for 
$q_{0} =0.5$, 
or 3.2 h$_{70}^{-1}$ kpc long for $q_{0} =0.1$.  In section 6, we 
examine in detail the question of whether or not O1 is real. In 
this section we discuss the properties of O1 assuming that it 
is real and is associated with the DLA absorber. 

The faintness and diffuse nature of object O1 make its photometry 
rather difficult. We estimated the flux from this object in the 
PSF-subtracted image, using three different procedures, and then took 
an average of the three values. 

Since accurate aperture photometry is 
difficult, we first estimated the flux by subtracting the PSF star 
from object O1, now multiplying the star by a factor large enough 
to make object O1 look indistinguishable from noise. This PSF 
multiplying factor can then be used directly to estimate the flux 
of object O1, since the PSF star P330E is a well-calibrated NICMOS 
photometric standard. From this method, we deduce that object O1 is 
$2.55 \times 10^{-4}$ as bright as the 
PSF star P330E. This implies a flux of 3.22 ADU s$^{-1}$ or 7.1 $\mu$Jy 
in the F160W filter. To convert the count rate to flux, we used 
the NICMOS photometric calibration factor of $2.190 \times 10^{-6}$ 
Jy/(ADU s$^{-1}$) for the F160W filter. This factor was derived 
using the solar-type photometric standard star P330E (that we also 
used for PSF subtraction). 

As a rough check of the above flux value, we also did aperture photometry  
on a circular aperture 4 pixels in diameter centered on O1 using the 
IRAF task apphot. A constant background value was subtracted as the 
sky value. (This constant, estimated as the average of the mean values per 
pixel of about 20 10x10 subregions in different parts of the image,  
had a very low mean 
value and therefore made negligible change to the final flux values.) 
This yields 2.55 ADU s$^{-1}$, i.e., 5.6 $\mu$Jy 
before correcting for aperture effects. For 
reference, the 1 $\sigma$ noise level in the PSF-subtracted image is 
about 0.13 ADU s$^{-1}$ per pixel (0.28 $\mu$Jy per pixel) 
in a circular annulus 0.2 $\arcsec$ wide centered at 0.3 $\arcsec$ 
from the quasar center. The corresponding noise levels at 
0.5 $\arcsec$, 0.7 $\arcsec$, 0.9 $\arcsec$, and 
1.1 $\arcsec$ from the quasar center are 
0.023, 0.014, 0.012, and 0.012 ADU s$^{-1}$ per pixel 
(i.e., 0.051, 0.031, 0.027, and 0.026 $\mu$Jy 
per pixel respectively). 

The 4-pixel diameter circular aperture covers most of  
the region of emission in O1 and avoids the residuals near the 
center of the quasar and very narrow features that we think arise 
from residual PSF differences. 
This region however does not include the pixels at the extreme ends of 
the ``major axis'' of O1, which is about 6 pixels long. We therefore 
also estimated the flux by doing a pixel-by-pixel addition over the 
region actually occupied by O1, which gives 2.63 ADU s$^{-1}$, i.e., 
5.8 $\mu$Jy. The flux values estimated by both the aperture photometry 
methods need to be 
corrected for the fact that a significant fraction of the energy of 
even a point source lies outside the radius of 2 pixels. Using 
aperture photometry on the standard star P330E, we estimate that the 
aperture correction factor between radii $r=2$ and $r=7.5$ pixels is 
1.625. A further factor of 1.152 has been estimated for camera 2 filter 
F160W for the aperture correction from a 7.5-pixel radius aperture to the 
total flux, based on standard NICMOS photometric calibrations made with 
the standard star P330E. Thus, the total aperture correction factor 
is 1.872 for the second method. We note, however, that there is a 
roughly 10 $\%$ uncertainty in the aperture correction factor. Schneider 
et al. have estimated the above correction factor to be 2.08. Taking an 
average, we adopt an aperture correction factor of 1.98 $\pm 0.1$. 
Since the region used in the third 
method is approximately also 4 pixels in diameter (although slightly 
bigger near the ends of the ``major axis'' of O1), 
the aperture correction is (at least) 1.98 in this case. We therefore 
use this factor for the third method of flux estimation also, although 
it is hard to be sure of the exact aperture correction in this case. 
After the aperture corrections, we derive flux values of 11.0 $\mu$Jy 
with the second method and 11.4 $\mu$Jy with the third method. 

On averaging the three flux values derived above, 
we estimate a flux of $9.8 \pm 2.4$ $\mu$Jy for the flux from object O1.  
This corresponds to $m_{F160W} = 20.11_{-0.24}^{+0.30}$ 
(taking the zero magnitude  
to correspond to 1083 Jy in the Johnson system). Here we have used 
equal weights for the three values while averaging, although we 
note that the value obtained by subtracting a point source is likely 
to be more accurate than the other two values. 
Note that the error estimate indicates the standard deviation among 
the three flux estimates obtained by the three methods, and thus 
reflects the uncertainties in the size and shape of object O1. 
For comparison, the 1 $\sigma$ uncertainty in the background near O1 
is 0.051 $\mu$Jy per pixel, or $\approx 0.2$ $\mu$Jy over the region of 
$\approx 12$ pixels occupied by O1.

The observed F160W flux corresponds to a luminosity (at mean rest 
frame wavelength of 0.55 $\mu$m) of about $1.5 \times 10^{10}$ 
h$_{70}^{-2}$ L$_{\odot}$ for $q_{0} =0.5$ and about 
$2.8 \times 10^{10}$ h$_{70}^{-2}$ L$_{\odot}$ for $q_{0} =0.1$. 
Thus, object O1 is fainter than an L$_{*}$ galaxy at $z=1.89$ by 
0.2-0.9 magnitudes. If O1 is not the DLA, the DLA is even fainter. 
Our results here are consistent with those of Djorgovski (1997), 
who reported  a possible counterpart to the $z=4.10$ DLA toward DMS 
2247-0209. That DLA candidate is located $\approx 3.3 \arcsec$ 
from the quasar (i.e. 22 h$_{70}^{-1}$ kpc for q$_{0}=0.1$), with 
an inferred continuum luminosity of 0.5 L$_{*}$. 

\subsection{NON-CORONAGRAPHIC F190N IMAGES}

At a redshift of $z_{DLA}=1.892$, any H-$\alpha$ emission would be 
expected to lie at $\lambda_{obs}=1.898$ $\mu$m, which is very close 
to the center and mean $\lambda$ of 1.900 $\mu$m for the filter 
F190N. Thus, the narrow-band images in filter F190N are expected to 
reveal any redshifted H-$\alpha$ emission from the DLA. Fig. 5b 
shows the PSF-subtracted F190N image using the PSF image of the star 
P330E observed on May 29, 1998. The residual image shows an emission 
feature in roughly the same place ($\approx 0.28$ $\arcsec$  
away from the quasar center to the lower right) and with roughly the 
same size (0.42 $\arcsec$ long) as the feature seen in the 
non-coronagraphic F160W image. This feature is more clearly evident 
if the images are sub-sampled by a factor of 2 (as discussed in 
sec. 6.7 and Fig. 16). As in the F160W image, this feature also 
does not disappear after realigning or rescaling the PSF. This suggests 
that feature O1 may be a real object. If O1 is associated with the 
DLA absorber at $z=1.892$, 
the absorber is $\approx$ 2.5 h$_{70}^{-1}$ kpc long for 
$q_{0} =0.5$, or $\approx$ 3.4 h$_{70}^{-1}$ kpc long for $q_{0} =0.1$.

As in the case of the broad-band images, the photometry of O1 
is rather difficult. We do it in three different ways and take an 
average. In an attempt to get a flux  estimate free of the uncertain 
aperture correction factor, we first subtracted the standard star 
P330E from O1, scaling the star such that the feature O1 just 
disappears. This method gives a flux in O1 of $4.11 \times 10^{-4}$ 
times that of P330E. This corresponds to a flux of 0.204 ADU s$^{-1}$, 
i.e. 9.1 $\mu$Jy.  Here we have used the NICMOS photometric 
calibration factor of $4.455 \times 10^{-5}$ Jy/(ADU s$^{-1}$) 
for the F190N filter. Aperture photometry yields a flux of 
0.114 ADU s$^{-1}$ in a 4-pixel diameter circular aperture centered on 
the center of O1. A pixel-by-pixel addition over the region occupied 
by O1 gives a flux of 0.132 ADU s$^{-1}$. Based on the photometry of 
the standard star P330E, we estimate that the aperture correction factor 
between $r=2$ and $r=7.5$ is 1.691. The aperture correction factor 
between $r=7.5$ and the total flux is expected to be 1.159 for the 
F190N filter. This implies a total aperture correction factor of 1.960 
for the $r=2$ values.  But a $\approx 10 \%$ uncertainty exists 
in the aperture correction factor, similar to that discussed above for 
the F160W images. We therefore adopt an average aperture correction 
factor of 2.06. Applying this aperture correction, we get flux 
values of 0.234 ADU s$^{-1}$ (10.4 $\mu$Jy) and 0.272 ADU s$^{-1}$ 
(12.2 $\mu$Jy), respectively for the second and third methods. 
Averaging the three flux values 
obtained by the three methods, we get $10.6 \pm 1.5$ $\mu$Jy. The 
error bar of 1.5 $\mu$Jy denotes only the standard deviation among the 
three values and thus reflects the uncertainties arising from the 
lack of knowledge about the size and shape of O1. For comparison, 
the 1$\sigma$ noise levels in the F190N image (after PSF subtraction) 
at $r=0.5$ $\arcsec$, 0.7 $\arcsec$, 0.9 $\arcsec$, and 1.1 $\arcsec$ 
from the quasar are 0.0027, 0.0019, 0.0018, and 0.0016 
ADU s$^{-1}$ per pixel,  i.e., 0.119, 0.084, 0.081, and 0.073 $\mu$Jy 
per pixel, respectively. Thus the 1 $\sigma$ sky noise uncertainty in 
the total summed flux over 
the $\approx 12$ pixel region occupied by O1 is $\approx $ 0.4 $\mu$Jy 
(using the noise estimates just outside O1 at $r=0.5$ \arcsec). 

The expected F190N continuum must be subtracted from the observed 
flux in order to determine if a statistically significant redshifted 
H-$\alpha$ excess exists. We estimate the continuum under the F190N 
filter by scaling the F160W image using the relative photometric 
calibration of the two filters. 
We find that, in fact, this expected continuum flux agrees almost 
completely with the observed F190N flux. After subtraction of the 
expected F190N continuum image (scaled from the F160W image) 
from the observed F190N image, we find a very marginal 
excess of 0.0074 ADU s$^{-1}$. With the aperture correction, this  
corresponds to 0.68 $\mu$Jy. The 1 $\sigma$ noise level in the 
F190N-F160W image is 0.0026 ADU s$^{-1}$ per pixel just outside the 
location of O1. This noise level corresponds to a 1 $\sigma$ 
uncertainty of 0.4 $\mu$Jy in the total flux summed over the region 
occupied by O1. The slight excess at the location of 
O1 in the F190N-F160W image is thus not statistically significant.  
We therefore conclude that the contribution to the F190N flux 
from redshifted H-$\alpha$ emission is negligible. 
It is not likely that we could have missed the H-$\alpha$ 
emission from O1. The H-$\alpha$ emission from the DLA could lie 
outside the F190N bandpass only if the DLA galaxy is lower in 
velocity by more than 980 km s$^{-1}$ or higher in velocity by 
more than 1770 km s$^{-1}$ 
from the absorption redshift. Such offsets are higher than the 
observed internal velocity dispersion in any typical single galaxy. 

Integrating over the FWHM of the F190N filter, assuming no dust 
extinction, and using the prescription of Kennicutt (1983) for 
conversion of H$\alpha$ luminosity to SFR, the nominal marginal 
excess of 0.68 $\mu$Jy in the F190N-F160W image corresponds to 
an SFR of 1.1  $h_{70}^{-2}$ M$_{\odot}$ yr$^{-1}$ for $q_{0} =0.5$, 
or 2.0 $h_{70}^{-2}$ M$_{\odot}$ yr$^{-1}$ for $q_{0} =0.1$. To derive 
a better estimate of the uncertainty in the H-$\alpha$ flux, 
we experimented with subtractions of the PSF-subtracted F190N and 
F160W images. In a 4-pixel region (roughly the size of our resolution 
element), an H-$\alpha$ emission strength of about 0.016 ADU s$^{-1}$ 
(0.71 $\mu$Jy) would yield S/N = 3. With an aperture 
correction factor of 3.41, this corresponds to a total 3 $\sigma$ 
flux limit of 0.054 ADU s$^{-1}$ or 2.4 $\mu$Jy. This translates into  
a 3 $\sigma$ upper limit on the SFR of 4.0 $h_{70}^{-2}$ 
M$_{\odot}$ yr$^{-1}$ for 
$q_{0} =0.5$ or 7.4 $h_{70}^{-2}$ M$_{\odot}$ yr$^{-1}$ for 
$q_{0} =0.1$. (We consider 
the possibility of dust extinction in section 7.3 below.)   

\subsection{CORONAGRAPHIC F160W IMAGES}

An F160W coronagraphic image of the quasar is shown in Fig. 6a, 
in which the coronagraphic hole is masked out. Almost 
all the flux seen in this reduced coronagraphic image is due 
to residual  
scattered light from the quasar, and ``glints'' from the 
edge of the hole. After subtraction of a reference PSF image 
using observations of the star GL83.1, these artifacts 
disappear almost entirely (Fig. 6b).  The bright emission 
feature about 0.25 $\arcsec$ to the lower right of the quasar center, 
seen in Figs. 4b and 5b, is just inside the coronagraphic hole 
and is therefore not seen in Fig. 6b. However, the coronagraph 
is very effective in reducing the quasar light outside of the 
coronagraphic hole, and can therefore be used to look at other 
objects in the field. 

A weak feature (which we name object ``O2'') remains after 
PSF subtraction (to the ``top left'' of the hole, about 
$0.7$ $\arcsec$  away from the quasar center). This 
feature is dominated by four knots of continuum emission. 
No artifacts resembling this feature have been seen in the 
coronagraphic images of PSF stars from other NICMOS GTO programs. 
By contrast, the knots seen to 
the ``lower left'' are known artifacts in the coronagraphic PSF. 
``O2'' is detected in the same 
place if the data for each of the two orbits are analyzed separately,  
which suggests that it may be real. It is not likely to be 
a trail of a cosmic ray event, since it is present in the images over 
a period of two orbits (much longer than typical time-scales 
for the decay of cosmic ray persistence in the NICMOS detectors). 
The knots in feature  
O2 are much weaker than the peak in feature O1, but are  
about 2-3 times the rms noise in the background. O2 has a 
total linear size of about 9-10 pixels, i.e. about 0.7-0.8 
$\arcsec$ . The 1 $\sigma$ noise levels per pixel in the PSF 
subtracted image at 0.3 $\arcsec$, 0.5 $\arcsec$, 
0.7 $\arcsec$, 0.9 $\arcsec$, and 1.1 $\arcsec$ from the quasar 
center are 0.032 ADU s$^{-1}$, 0.012 ADU s$^{-1}$, 
0.011 ADU s$^{-1}$, 0.0088 ADU s$^{-1}$, 
and 0.0094 ADU s$^{-1}$, i.e., 0.069, 0.027, 0.024, 0.019, and 0.021 
$\mu$Jy per pixel, respectively. Compared to the non-coronagraphic 
F160W image, these noise levels indicate factors of 4.06, 1.88, 1.30, 
1.41, and 1.26 improvements, respectively, in the 1 $\sigma$ 
sensitivities at 0.3 $\arcsec$, 0.5 $\arcsec$, 0.7 $\arcsec$, 
0.9 $\arcsec$, and 1.1 $\arcsec$ 
from the quasar center. These factors are much smaller than 
those typically reported for NICMOS coronagraphic performance, 
because the low signal from our faint quasar makes our 
observations read noise dominated.

The results of coronagraphic imaging (e.g. appearance of object O2) 
are not expected to be very sensitive to the data reduction procedures. 
No dithering was used between the coronagraphic exposures, to ensure 
that the quasar always remained in the coronagraphic hole. Therefore 
the individual coronagraphic exposures were not registered 
before they were combined. The quasar was acquired with 
on-board target acquisition and placed in the coronagraphic hole at 
the beginning of the first orbit. The quasar was placed in the same 
position in the second orbit. Guide star acquisition was done at the 
beginning of each of the two orbits using the same guide stars. 
Therefore we believe that there are no  
significant offsets between the quasar's positions in the hole in 
the various coronagraphic exposures. Indeed, as mentioned 
above, the coronagraphic images obtained in each orbit separately 
show the object O2, which appears similar in both the orbits. The fact 
that features in the coronagraphic PSF other than object O2 disappear 
after the PSF subtraction also suggests that object O2 is not the 
result of misregistration of the individual exposures. We therefore 
believe that object O2 is likely to be real. 

This feature O2 is detected marginally in the non-coronagraphic 
F160W image (Fig. 4b) due to the higher scattered light from 
the quasar in that image. We note that, while of lower 
sensitivity, the faint compact emission features to 
the top left of the quasar in this image are at positions 
similar to those of the O2 knots in the coronagraphic image.   
Object O2 is not seen in the narrow-band image in Fig. 5b. 
However, considering that it is much  fainter than object O1,  
it is not entirely surprising that any emission from O2 is not 
detected in the narrow-band images (which are about 10 times 
less sensitive, at the separation of O2 from the quasar). 
Considering this, and its faintness and larger angular distance 
from the quasar compared to O1, it is not completely clear 
whether the feature O2 has any connection with the DLA.  But 
it may be associated with the DLA or its companions. It is 
also possible that objects O1 and O2 are associated with the 
host galaxy of the quasar rather than the DLA. We discuss this 
possibility further in section 7.4. If O2 is indeed associated 
with the DLA at $z=1.892$, then it has a size of 4-5 
h$_{70}^{-1}$ kpc for $q_{0} = 0.5$. 

\section{IS OBJECT O1 REAL?}

In view of the low S/N of object O1 and its small angular separation from 
the quasar in our non-coronagraphic 
F160W and F190N images, we carried out a number of tests 
on the images to investigate whether O1 is real or merely 
an artifact of the data reduction or PSF subtraction procedures. 
Here we describe these tests, listing the potential sources of 
error that we investigated in each case, and the corresponding 
results. 

\subsection{Is minimum variance the right criterion in PSF subtraction?}

We have registered and normalized the PSF star by varying 
the position and multiplicative scaling factor of the PSF star so as 
to minimize the variance in the region of 
interest near the quasar. This seems to be the most 
objective way of judging the goodness-of-fit of the PSF subtraction. 
To determine whether any bias could be 
caused by the use of the minimum-variance criterion, 
we also verified that the results from this method are closely 
consistent with K. McLeod's method of
forcing the intensity at the first Airy minimum to zero (see McLeod, 
Rieke, \& Storrie-Lombardi 1999). The PSF star position given by 
the two methods for the optimum PSF subtractions in each 
case agree to within 0.002 pixels. The PSF normalization factors 
from the two methods agree to within about 2 $\%$. 
In either case, our broad conclusions about the nature of 
the residuals after PSF subtraction (including feature O1) are the 
same for both the methods. Therefore we believe 
that our strategy of minimizing the variance is sound. 

\subsection{Telescope breathing effects?}

To examine how sensitive the detection of the main 
emission knot O1 is to the fine structure of the PSF subtracted, we 
created a suite of difference images 
using a variety of reference PSFs for the 
non-coronagraphic broad-band and narrow-band images. 
We particularly sought to investigate the effects of HST 
breathing focus changes on our results. The changes in HST 
focus consist of two components. First, there is a long-term 
slow change caused by shrinkage in the 
Optical Telescope Assembly (OTA) of HST due to moisture desorption, 
which is periodically corrected by secondary mirror moves. In 
addition, short-term focus variations on the time scale of the HST 
orbit, arising from thermally driven displacements of the OTA 
secondary, can be even larger in magnitude than desorption correction 
compensations. 

Fig. 7 shows the effect of using various observations of 
the PSF star P330E on the non-coronagraphic F160W images. 
Fig. 7a is the same as Fig. 4b, while Figs. 7b, 7c, and 7d show 
the results obtained by subtracting images of P330E taken 
on different dates and with different breathing values.   
The breathing values denote the position of the HST 
secondary mirror in units of $\mu$m with respect to a 
common reference, i.e., with respect to the best focus of 
WFPC2 planetary camera (see Hershey \& Mitchell 1998). All four 
panels of Fig. 7 show the 
feature O1 in roughly the same place with other 
variations being much smaller in amplitude than O1. 
Fig 8 shows the difference of the PSFs used in making 
Fig. 7.  Figs. 8a, 8b, 8c show, respectively, (PSF for 
fig. 7a - PSF for fig. 7b), (PSF for fig. 7a - PSF for 
fig. 7c), and (PSF for fig. 7a - PSF for fig. 7d). The 
differences among the residuals in the different panels of 
Fig. 8 arise partly from 
breathing variations. But we note that the different dates 
for the reference PSF observations imply the use of 
different guide stars, and hence the PSF star would have 
landed on different pixels in these different images. 
Therefore the intra-pixel response function, in addition 
to focus changes, could also account for some of the 
differences between the different PSF images. {\footnote
{We further note that the breathing models of Hershey \& Mitchell  
(1998) have some uncertainty. This could give rise to 
some residuals in our difference images arising from differential 
inaccuracies in the breathing values for the 
quasar and the PSF star images predicted by the models.}}
In any case, the symmetric nature of the residuals in Fig. 
8 and the absence of the knot O1 in these images 
suggests that the latter feature is present in the quasar images, 
and not an artifact in any individual PSF image. 

Fig. 9 illustrates the effect of telescope breathing focus 
variations on the F190N images, with three different 
observations of PSF star P330E. Fig. 9a is the same as Fig. 
5b. The corresponding PSF star differences are shown in 
Fig. 10. Figs. 10a and 10b show, respectively, (PSF for 
Fig. 9a - PSF for Fig. 9b), and (PSF for Fig. 9a - PSF for 
Fig. 9c). The feature O1 is detected in the same place in 
all the panels of Fig. 9, and most of it is not seen in 
the PSF star differences (Fig. 10). 

\subsection{Using different PSF stars: Color mismatch 
between quasar and PSF star?}

Color terms in the PSFs are potentially important sources 
of error in the difference images. For our primary PSF subtractions 
we have used the solar analogue P330E ($m_{F110w} - m_{F160W}=0.44$, 
$m_{F160W} - m_{F222M}=0.08$) as described in the previous section. 
However, we also experimented with a red PSF star BRI0021 
($m_{F110w} - m_{F160W}=1.17$, $m_{F160W} - m_{F222M}=0.80$). 
Fig. 11 shows the effect of using four different PSF stars 
(P330E, BRI 0021, Q1718PSF, and GSC4) in the top left, top right, 
bottom left, and bottom right panels, respectively. Note that the 
breathing values for the four observations are quite different, 
which could explain the differences in the appearance of O1. In 
any case, all the images show an excess residual at the location of 
O1, while such excesses are not seen in the differences of the PSFs 
themselves (shown in Fig. 12). Thus object O1 is probably not an 
artifact caused by color mismatch between the quasar and the PSF star. 

\subsection{Calibration defects for column 127 influencing the 
centroids?} 

Two of our five dither positions for the non-coronagraphic F160W 
images had the quasar image near column 127 
(in camera 2 detector coordinates). This column is 
well-known to be ``photometrically challenged''. 
A ``bad stripe'' in this column results if the dark frame 
used for the calibration is not a perfect match to 
the dark current in the actual observations. We corrected  
for the ``bad stripe'' in this column by including it in the 
bad-pixel mask used while imcombining the five dither positions. 
However, potentially this column may influence the 
centroids of the images at the two dither positions and hence the 
centroid of the imcombined image. To explore this possibility, we 
looked at each of the three remaining dither positions not affected 
by column 127. Fig. 13 shows central regions of the PSF-subtracted 
images for these three individual dither positions in top left, 
top right, and bottom left panels. The bottom right panel shows the 
result of PSF subtraction for the image obtained by combining 
only these three dither positions. For obtaining the minimum-variance 
solutions for these PSF subtractions, we have excluded the PSF cores 
while determining the variance. The first dither position as well as 
the combined image (bottom right panel) show an asymmetric 
excess emission near the position of object O1. This 
suggests that the feature O1 is not caused by errors 
arising from column 127, since none of the dither 
positions considered in Fig. 13 include this column. 

\subsection{Persistence effects from the quasar image at 
previous dither positions?} 

The experiments with PSF subtraction on the individual 
dithers described in test (6.4) above also help to show 
that image persistence effects are not important in 
causing feature O1. This is because even the very 
first dither position (which should not suffer from quasar 
persistence effects) shows the presence of an asymmetric feature 
at the location of O1 (see the top left panel of Fig. 13). 
Also, the quasar LBQS 1210+1731 is faint, so it is not likely 
to cause persistence effect. The fact that O1 has roughly equal 
intensity in all dither positions and does not go away even in the 
final dither position implies that O1 is also not a left over 
persistence image from a bright object or cosmic ray detected 
before the start of our observations. 

\subsection{Difference between ``camera 1-2 focus'' vs. 
``camera 2 focus''?}

Our quasar images were obtained with the NICMOS 
internal focusing mechanism optimized for parallel 
camera 1 and 2 operations, whereas all of our reference 
PSFs were taken at the camera 2 exclusive focus. A very slight 
deviation from confocality in the two 
cameras results in a wavefront error of 0.049 $\mu$m 
mm$^{-1}$ of focal dispersion. The ``focus error'' in 
camera 2 at the critically sampled $\lambda$ (1.75 $\mu$m) 
is $\lambda/33$ with the focus at the common position. 
While small, this ``focus error'' can affect the fine 
structure of the PSF to a very small degree, as higher 
order aberrations also change (with an aggregate 
power of about half the focus error). 

To investigate whether this effect can give rise to feature O1, we took 
two approaches. In the first 
approach, we used images of a star actually observed at 
the compromise focus ``camera 1-2''. There were no 
systematic PSF star measurements made at this focus 
position during the NICMOS calibration program. However, 
we found a star in one of our images of the galaxy 
cluster CL0939+47 taken for another NICMOS GTO program. 
Fig. 14a shows the PSF subtraction results obtained for 
our quasar image using this 
observed PSF at the compromise focus. The fact that 
some of the features in O1 are still seen while some disappear 
suggests that some of the O1 features (e.g. 
the blob to the right of the core) could be real. 

In the second approach, we constructed simulated NICMOS camera 2 
PSFs using the Tiny Tim program (version 4.4, 
Krist \& Hook 1997). We constructed simulated PSFs for 
the two PAM (pupil alignment mirror) positions 
corresponding to the two focii at the time of our quasar 
observations on July 22, 1998 and the P330E observations 
on July 8, 1998. These two simulated Tiny Tim PSFs differ 
only in this focus position and both used the same 
values of rms jitter (0.007 $\arcsec$), same x and y 
pixel positions for placement of PSF star center, same 
pixel size, etc. After making these two simulated PSFs, 
we corrected them for the slight relative difference in 
the actual x and y plate scales (interpolated in time 
for the dates of the observations for LBQS 1210+1731
 and the observations for P330E). This slight 
repixelization corrects for the fact that Tiny Tim 
creates images with equal x and y pixel scales, whereas 
the actual x and y pixel scales differ by $0.9 \%$. The 
difference of the two simulated Tiny Tim PSFs corrected 
for the unequal x and y pixel scales is shown in Fig. 14b. 
The difference does show some residuals along the diagonal 
directions. However, these are symmetric in shape on both 
sides of the center, and are accompanied by much larger residuals 
in the core of the image. Fig. 14c shows, on the same stretch as 
Fig. 14a, the difference of the two simulated Tiny Tim PSFs 
after normalizing each to match the quasar. The residuals in 
Fig. 14c are much weaker than object O1.  
A simple relative translation between the images for the two 
focii cannot give rise to a feature as significant 
as O1 without causing a much larger residual in the core. 
We therefore conclude that while the difference in the 
PAM positions for the quasar and the PSF star could cause some 
of the residuals in our PSF subtractions, they cannot be 
the major source of these residuals. 

To pursue this analysis further, we took the ratio of 
the two simulated Tiny Tim PSFs after correction for 
plate-scales {\footnote{Here, by the ratio of the 
simulated Tiny Tim PSFs, we mean the ratio of the PSF 
with the PAM position for LBQS 1210+1731 to the PSF with 
the PAM position for the star P330E, the same two PSFs 
whose difference is shown Figs. 14b and 14c.}}, and multiplied this 
ratio by the actual 
observed P330E PSF to make our ``best-guess'' PSF.  
The resultant PSF has the advantages of combining the 
correct focus (PAM) position (because of the Tiny Tim simulation), 
the best estimate of breathing (because 
of use of the actual observation of P330E which matches closely in 
breathing with the LBQS 1210+1731 data), 
and any other actual optical effects that Tiny Tim does 
not simulate adequately. In the bottom right panel of 
Fig. 14, we show the resultant image obtained after subtracting 
this ``best-guess'' synthetic P330E PSF from 
the LBQS 1210+1731 data. Once again, excess emission is 
seen at the position of O1. This suggests that O1 is not caused by 
artifacts of relative focus difference 
(``camera 1-2'' focus vs. ``camera 2'' focus) between the quasar and 
the PSF star.

\subsection{Errors in imcombining or interpolating the dithers?}
 
We used Nicred 1.8 to interpolate the registered dithers 
onto a grid of single camera 2 pixels, or onto a grid 
of half integer camera 2 pixels. This magnification (repixelization) 
or lack thereof made  
little difference in the result. This is clear from 
Fig. 15, which shows the F160W images made on using 
grids of single camera 2 pixels (Fig. 15a), 
and half integer camera 2 pixels (Fig. 15b). Figs. 
15c and 15d show the same figures with pixel 
replication instead of cubic convolution interpolation 
in the IDP3 display. The similarity between the left 
and right panels is reassuring and results from camera 
2 being nearly critically sampled at 1.6 $\mu$m. The 
same was also found to be true for the F190N images. 
Fig. 16 shows the F190N images made on using 
grids of single camera 2 pixels and  
half integer camera 2 pixels (Figs. 16 a and 16b shown with 
cubic convolution and Figs. 16c and 16d shown with pixel 
replication). Both Figs. 15 and 16 suggest that object O1 
is likely to be real and does not arise from interpolation errors. 

\subsection{Effects of asymmetries or saturation in the core?}

Based on our experience with other NICMOS GTO data, the 
core of the PSF often shows paired positive and negative residuals 
after PSF subtraction. In case the core of the quasar and PSF star 
images have some asymmetries which 
might mimic an O1-like feature after PSF subtraction, 
we have also done the PSF subtraction without including 
the core for the variance calculation. The right panel in 
Fig. 17 shows the non-coronagraphic F160W image obtained after 
minimizing the variance in the PSF subtraction without including 
the core of the image for variance calculation. The region 
thus excluded, shown with a circular mask, covers the 
region up to the first Airy minimum. The left panel in 
Fig. 17 shows the non-coronagraphic F160W image obtained when the 
core is included (same as the image shown in Fig. 4b, except that 
the central core is masked in the display only for easy comparison 
with the right panel of Fig. 17). The similarity between the 
two panels of Fig. 17 (presence of a feature at the 
position of O1) suggests that asymmetries in the 
core are not the source of O1. 

This same experiment also shows that saturation in the 
core of the quasar image cannot be a major problem (since 
the results obtained by including or excluding the core 
agree very closely). The individual 512-s exposures at 
each of the 5 dither positions in the F160W image of our quasar 
have peaks of about 15000 ADU or 83000 e$^{-}$ in 
the quasar central pixel. Thus we do expect that they 
should not be saturated, given the $98 \%$ linearity saturation 
limit of 173,000 e$^{-}$ for camera 2. 

\subsection{Errors in alignment of PSF star with respect to 
the quasar?}

Fig. 18 shows the effects of shifting the PSF star by 0.1 
pixel in various directions relative to the quasar on the 
difference F160W images. Fig. 18e is the optimum 
minimum-variance solution (same as Fig. 4b). Figs. 18b, 
18h, 18d, and 18f show the PSF subtractions obtained 
after shifting the PSF star by 0.1 pixel in the top, bottom,
left, and right directions, respectively, with respect 
to the quasar. Figs. 18a, 18c, 18g and 18i show the 
corresponding results on shifting the PSF star by 0.1 pixel 
in the upper left, upper right, lower left and lower 
right directions, respectively, with respect to the 
quasar. The large residuals in the core caused by even 
the slight shifts illustrate that the PSF star is very well 
aligned with respect to the quasar in the optimum PSF 
subtraction (Fig. 18e). We have shown shifts of 0.1 
pixel in Fig. 18 to make the changes easier to view. 
But, judging by the minimum in the variance, we believe 
that our relative alignment of the quasar and PSF star images 
is good to at least 0.01 pixel. 
Thus, errors in alignment of the PSF star with respect to 
the quasar should be insignificant.

\subsection{Errors in stacking the individual dither positions?} 

The different dither positions are registered in Nicred 1.8 using 
cross-correlation. To check the accuracy of the image registration, 
we compared the centroids of the images at the various dither positions 
after registration. The 1 $\sigma$ variation among the centroid values 
of the different dither images was found to be about 0.04-0.06 pixels, 
for both the quasar and the PSF star. 
The centroid values for any individual dither position calculated 
from different methods (tasks imexam, center, and starfind in 
IRAF) were also found to agree within about 0.06 pixels. Thus, 
there is a small uncertainty in the centroid values, but it 
does not seem large enough to cause a feature such as O1.  
The fact that the individual dither positions show some excess at 
the position of O1 (Fig. 13) also suggests that O1 is not a 
spurious feature resulting from stacking errors. 

\subsection{Comparison with other data}

(a) Comparison of PSF stars with each other: 
PSF stars seem to subtract very well from each other, with 
the same caveat about color and breathing.  There is no 
hint of a $1\%$ residual at the position corresponding to 
the feature O1 (See fig. 12.)

(b) Same reduction on other quasar data: We reduced the 
data for quasar Q1718+4807 at z=1.084 (a quasar without 
a DLA absorber) from another NICMOS GTO program, using 
the same reduction and PSF subtraction procedures as we 
have used for LBQS 1210+1731. We do not see the object 
O1 there.  

We have also reduced the data for the other quasars with 
DLAs from our sample, which will be described in separate papers 
(Kulkarni et al. 2000b, 2000c). Comparing the results for 
LBQS 1210+1731 with the results for those quasars, we find that 
some of the residuals in the PSF subtractions appear similar, 
while some of the features are different. This suggests that part 
of the emission at the position of O1 is likely to be real, 
although part of it could be some artifact that we have not yet 
understood despite the large number of data analysis experiments 
described above.

\subsection{Summary of Results from Various Data Analysis Tests}

Overall, we conclude that the best-fitting PSF and several others 
with reasonably close breathing values suggest a possible detection 
of an object (object ``O1'') located at about 0.25 $\arcsec$  from 
the quasar center, in both the F160W and F190N images. The appearance 
and properties of this object are more sensitive to the important step 
of PSF subtraction than to other data reduction steps such as flat 
fielding. However, our extensive tests suggest that this object is not 
an artifact of color or focus mismatch or spatial misalignment between 
the quasar and PSF star images. It is also not caused by image 
persistence or saturation or by the procedures used for interpolation 
or stacking of the individual images. We therefore believe that 
object O1 is likely to be real. 

The most relevant broad and narrow band summary images showing 
object O1 are the zoomed, magnified, PSF subtracted images in 
Figs. 15 (b) and 16 (b).
The small angular separation of O1 from the quasar suggests that 
it is likely to be associated with the DLA absorber. The corresponding 
impact parameter is 1.5 h$_{70}^{-1}$ kpc for q$_{0}$ = 0.5 or 2.0  
h$_{70}^{-1}$ kpc for q$_{0}$ = 0.1. We have thus probed regions far 
closer to the quasar sight-line than most previous studies of 
high-redshift intervening DLAs. Object O1 marks the closest detected high-redshift DLA candidate yet to any 
quasar sight line. Object O1 is 0.4$\arcsec$ long. If O1 is the DLA 
at $z=1.89$, this translates into 2.4 h$_{70}^{-1}$ kpc for q$_{0}$ = 
0.5 or 3.2 h$_{70}^{-1}$ kpc for q$_{0}$ = 0.1. It has a luminosity 
(at mean rest frame wavelength of 0.55 $\mu$m) of about 
$1.5 \times 10^{10}$ h$_{70}^{-2}$ L$_{\odot}$ for $q_{0} =0.5$ and about 
$2.8 \times 10^{10}$ h$_{70}^{-2}$ L$_{\odot}$ for $q_{0} =0.1$. Obejct 
O1 is thus fainter than an L$_{*}$ galaxy at $z = 1.89$ by 0.2-0.9 
magnitudes. The comparison of the broad and 
narrow band fluxes implies a nominal statistically insignificant 
SFR of 1.1 h$_{70}^{-2}$ $M_{\odot}$ yr$^{-1}$, with a 3 $\sigma$ upper 
limit of 4.0 h$_{70}^{-2}$ $M_{\odot}$ yr$^{-1}$, for q$_{0}$ = 0.5.

Another fainter object O2 which consists of 4 knots of continuum 
emission is also seen in our images. (See Fig. 6b.) This object, at 
angular separation of 0.65 $\arcsec$ from the quasar (well outside the 
first Airy ring of the quasar PSF) is also not a known artifact of the 
PSF. It is thus also likely to be real and may be a companion to the 
DLA. The spatial extent of O2 is 4-5 
h$_{70}^{-1}$ kpc and its projected impact parameter is 3.8 h$_{70}^{-1}$ 
kpc. Object O2, like object O1, is also closer to the quasar sightline 
than most other high-redshift DLA candidates detected before. 

We note, however, that because of the faintness 
and proximity of O1 to the quasar, we cannot completely rule 
out the possibility that this feature could partly be some as yet 
unknown artifact of the PSF (that is not simulated by Tiny Tim either). 
If any such errors are the actual cause of O1, then the DLA absorber 
and the quasar host galaxy are even fainter than O1. In that case, we 
can use our images to put very sensitive upper limits on the size 
and brightness of both the DLA absorber and the quasar 
host. We discuss the implications of our observations in the 
following section. 

\section{DISCUSSION}

The most important result from our observations is that 
there are no large bright galaxies close to the quasar 
in the field of the DLA absorber toward LBQS 1210+1731. Feature O1 is 
the most likely candidate for any object associated with the DLA. 
In subsections 7.1, 7.2, and 7.3, we assume that object O1 is 
associated with the DLA to derive constraints on various properties of 
DLAs. But we also consider alternative possibilities in subsection 7.4, 
mainly the possibility that O1 may be associated with the host 
galaxy of the quasar. 

\subsection{CONSTRAINTS ON SIZES AND MORPHOLOGY OF DLAs}

Our observations show no evidence for a big, well-formed galaxy as 
expected in some scenarios for the DLAs [e.g., the proto-spiral model 
suggested by Wolfe et al. (1986), Prochaska \& Wolfe (1997, 1998), 
Jedamzik \& Prochaska (1998)]. Feature O1 has an estimated size of 2-3 
$h_{70}^{-1}$ kpc, while feature O2, if real, consists of small 
knots spread over about 4-5 $h_{70}^{-1}$ kpc. 
Thus, these data suggest that the absorber is compact 
and clumpy, as expected in the hierarchical picture of 
galaxy formation. However, it is hard to be completely certain 
of the morphology, partly because of the 
sensitivity of the detailed image structure to the 
various factors discussed in sec. 6. Furthermore, it is possible that O1 
and O2 are the brightest regions within 
a bigger galaxy, the rest of which we cannot see. Thus, 
we cannot completely rule out the large disk scenario, although 
the compact sizes and low SFRs suggest that the hierarchical picture 
may be favored. Analysis of the other DLAs from our sample and further 
deeper observations will help to more definitively distinguish between 
the large disk vs. hierarchical models. 

\subsection{CONSTRAINTS ON ENVIRONMENT OF DLA ABSORBERS}

Apart from features O1 and O2 very close to the quasar, 
our images show two prominent galaxies in the non-coronagraphic 
F160W image (one in the upper left 
corner or west of the quasar and the other at the middle 
of the left edge of the image or roughly north of the quasar-- 
see Fig. 1). 
There is also a third very weak feature to the 
left (roughly north) of the quasar, a little less than 
half the way along the line joining the quasar and the 
galaxy at the middle left edge. The galaxy west of the 
quasar is just barely apparent in the non-coronagraphic 
F190N image, while the other two objects are not seen 
in the non-coronagraphic F190N image. The two prominent 
galaxies in the non-coronagraphic F160W image are off 
the field of the coronagraphic image, while the 
prominent galaxy seen at the bottom edge (northeast) of 
the coronagraphic F160W image is off the field of the 
non-coronagraphic images. It is possible that the faint 
feature to the left (north) of the quasar is spurious. 
But on running maximum-entropy and Lucy deconvolutions 
of the images, all the three objects (including the 
faint feature) in the F160W image were found to remain 
significant. These objects are likely to be galaxies in 
the same group as the DLA, although we do not have 
redshift information on them. In any case, they have 
fairly large impact parameters (4.52 $\arcsec$, 
11.00 $\arcsec$, and 10.96 $\arcsec$ for the faint 
feature, the galaxy to the west of the quasar, and the 
galaxy to the north of the quasar, respectively). At the 
redshift of the DLA absorber, these impact parameters 
would correspond to 26.8, 65.1 and 64.9 h$_{70}^{-1}$ kpc 
respectively, for q$_{0} = 0.5$. For q$_{0} = 0.1$, the 
corresponding values are 36.6, 89.0, and 88.6 h$_{70}^{-1}$ kpc. 
These large values make it unlikely for any of these features 
to be the DLA absorber itself. 
 
If real, the continuum emission knots in object O2 may 
be highlighting the brightest regions in a companion to 
the DLA galaxy. The roughly filamentary morphology may 
indicate an edge-on disk galaxy or a part of a spiral arm. 
Alternatively, it may suggest 
individual star-forming sub-galactic clumps 
formed in a filamentary over-dense region, similar to the 
filamentary arrangements of galaxies and sub-galactic 
units found in numerical simulations of structure 
formation. It is interesting to note that the WFPC2 observations of a 
$z = 2.811$ DLA by M{\o}ller \& Warren (1998) also indicate a 
filamentary arrangement of 3 bright objects, although on a much 
larger scale (separation of 21 $\arcsec$ ). The angular separation 
of their closest object from the quasar was 1.17 $\arcsec$, whereas 
for our features O1 and O2, the angular separations are 
$\approx 0.26 $\arcsec$ $ and $\approx 0.65 $\arcsec$ $, respectively. 
[We note, however, an 
important difference between our DLA and the DLA studied 
by M{\o}ller \& Warren. The latter has a redshift very 
close to that of the quasar ($z_{em, CIV} = 2.77$, 
$z_{em, [OIII]} = 2.788$, and $z_{em, H\alpha} = 2.806$).  
Therefore it is likely to be associated with the quasar 
and may not be representative of DLA galaxies in general.]

The 1 $\sigma$ noise levels far away from the quasar are 0.011 ADU 
s$^{-1}$ per pixel for our PSF-subtracted non-coronagraphic F160W 
image and 0.0088 ADU s$^{-1}$ per pixel for the PSF-subtracted 
coronagraphic F160W image. 
These levels translate into 0.024 $\mu$Jy 
per pixel and 0.019 $\mu$ Jy per pixel 
respectively. The corresponding 1 $\sigma$ noise 
equivalent magnitudes for the non-coronagraphic and coronagraphic 
F160W images are 26.6 magnitudes per 
pixel (21.0 magnitudes per square arcsecond) and 26.9 magnitudes 
per pixel (21.3 magnitudes per square 
arcsecond), respectively. For comparison, the Hubble 
Deep Field F160W images had a 1 $\sigma$ noise level of 
$1.22 \times 10^{-9}$ Jy per Camera 3 pixel (Thompson 
et al. 1999). Thus, for the field galaxies far from 
the quasar in the PSF-subtracted F160W observations, 
our images are about 5.1-5.4 magnitudes less deep than the 
Hubble Deep Field images. \footnote{Before doing the 
PSF subtraction, the 1 $\sigma$ noise levels 
far away from the quasar are $\approx 0.0046$ ADU s$^{-1}$ per 
pixel (0.010 $\mu$Jy per pixel or 27.6 mag 
per pixel) for the non-coronagraphic F160W image and 
$\approx 0.0084$ ADU s$^{-1}$ per pixel (0.018 $\mu$Jy 
per pixel or 26.9 mag per pixel) for the coronagraphic 
F160W image. The process of PSF subtraction decreases 
the 1 $\sigma$ deviations by a large factor near the 
quasar, but increases the noise far away from the 
quasar. This is because of the use of actually observed 
 PSF star images (with high but finite S/N) for PSF subtraction, 
which contribute to the noise level. But 
for reasons mentioned earlier, it is better to use 
observed PSF star images rather than Tiny Tim models 
to get good matches to the quasar PSF. The higher noise 
level in the coronagraphic F160W image before PSF 
subtraction compared to the non-coronagraphic F160W 
image seems to arise from the use of the target 
acquisition flat rather than the higher-S/N standard 
flat used for the non-coronagraphic image. In any 
case, our images both before and after PSF subtraction 
do not show any field galaxies other than those 
mentioned above.} 

Our images do not show any objects other than object O2 
in the close vicinity of the DLA. From the galaxy number 
count-magnitude relation based on deep NICMOS images (Yan et al. 1998), 
about 1 galaxy is expected for $H < 21$ in the camera 2 field. 
Thus our observations are consistent with these predictions within the 
uncertainties. There is no sign of strong clustering of galaxies 
around the DLA.  

\subsection{CONSTRAINTS ON STAR-FORMATION RATE AND DUST IN DLAs}

It is quite surprising, given the high sensitivity of our 
observations and the reasonably high rest-frame V-band luminosity 
of object O1, that O1 shows almost no detectable H-$\alpha$ 
emission. The lack of significant H-$\alpha$ emission in our images 
puts fairly tight constraints on the star-formation rate 
in the DLA toward LBQS 1210+1731, i.e. a $3 \sigma$ upper limit of 
$4.0$  h$_{70}^{-2}$ M${_\odot}$ yr$^{-1}$ for 
$q_{0} = 0.5$, if no dust is assumed. 
This is by far the most severe existing constraint on the SFR in 
high-$z$ DLAs. For comparison, the near-IR spectroscopic survey of 
Bunker et al. (1999), aimed at detecting H$\alpha$ from DLAs, gave  
typical upper limits of $\approx$ 15 M$_{\odot}$ yr$^{-1}$, for 
q$_{0}=0.5$ and $H_{0} = 70$ km s$^{-1}$ Mpc$^{-1}$. In Fig. 19, 
we compare the result from our data (shown as a filled triangle) 
with the $3 \sigma$ upper limits from Bunker et al. (1999) (shown as 
unfilled triangles). Our limit on the SFR marks an improvement by a 
factor of 3 over the tightest constraints of Bunker et al. (1999) 
on the SFR in DLA 
galaxies. The curve in Fig. 19 shows the predicted average SFR(z) in 
a DLA expected if DLAs are proto-disks, as derived by Bunker et al. 
(1999) using the closed-box model of Pei \& Fall (1995) for the global 
star formation rate. It is clear that our upper limit on the 
SFR is much lower than the predicted value at $z=1.89$. We note that 
the low SFR estimated here is consistent with the result of Djorgovski 
(1997) who reported SFR of $\approx 0.7$ M$_{\odot}$ yr$^{-1}$ in the 
$z=4.1$ DLA toward DMS 2247-0209, on the basis of a weak Ly-$\alpha$ 
emission line (assuming no dust extinction). (We note, however, that 
our limit is less sensitive to dust extinction uncertainties owing to 
the use of H-$\alpha$ rather than Ly-$\alpha$ emission.)

In principle, the lack of detectable H-$\alpha$ emission from the 
DLA could be because of dust extinction, in which case the actual 
SFR could be higher. In order to reconcile our upper limit of 
4.0 M$_{\odot}$ yr$^{-1}$ for q$_{0}=0.5$, $H_{0} = 70$ km s$^{-1}$ 
Mpc$^{-1}$, with the expectation of the closed-box proto-disk 
model of 38.6 M$_{\odot}$ yr$^{-1}$, an optical depth 
$\tau_{0.66 \mu m} \ge  2.3$ would be required at the rest frame 
H-$\alpha$ line, if a simple screen of dust in front of the DLA is 
assumed to extinguish the H-$\alpha$ emission. For extinction 
curves similar to those in the Milky Way, the Small Magellanic Cloud, 
or the Large Magellanic Clouds, this would imply $\tau_{B} \ge 3.4$ at 
$\lambda_{B} = 4400$ \AA. To have such high extinction, the DLA would 
be required to have a mean dust-to-gas ratio 
$k \equiv \tau_{B} (10^{21} / N_{HI}) \ge 8.7$. Even if the HI column 
density is assumed to be higher by a factor of $\sim 3$ at the position 
of O1 compared to the $N_{HI}$ detected in the DLA line (since the 
projected separation of O1 from the quasar would indicate that the DLA 
absorbing region may be a scale 
length away from the peak of emission from O1), one still requires   
a mean dust-to-gas ratio $k \gtrsim 3$. This is much higher  
than the mean dust-to-gas ratio of 0.8 for the Milky Way, or the 
typical value of $\sim$ 0.03-0.1 for the DLA galaxies, suggested by 
observations of background quasar reddening and heavy element depletions 
(see, e.g., Pei et al. 1991; Pettini et al. 1997 and references therein). 

It is, however, possible that the dust may be intermingled with the gas 
very close to the stars in the DLA. However, given the low dust-to-gas 
ratios seen in DLA absorbers, it is hard to imagine that most of the 
Ly-continuum  photons could be absorbed even before H-$\alpha$ photons 
can be produced. Thus, there is a good chance that the lack of 
H-$\alpha$ emission could be indeed because of low SFR. 

In the absence of dust obscuration, it follows from 
Fig. 19 that our results, as well as those of Bunker 
et al. (1999), indicate SFRs much lower than the 
expectations of the proto-disk model. This together with 
the compact sizes seen in our images again suggests 
that the observations do not agree with the proto-disk models. It is 
possible that O1 is a dwarf galaxy. Star formation in dwarf galaxies 
is inferred to proceed in 
bursts separated by quiescent periods lasting up to 
several Gyr (e.g., Grebel 1998). It may be that we 
are observing object O1 during such a relatively 
quiescent stage. It is also possible that O1 is a 
low-surface brightness galaxy, since such galaxies show 
lower SFR.  

\subsection{ALTERNATIVE POSSIBILITIES}

Finally, it is possible that object O1 is not the DLA absorber, but 
that it arises mostly in the quasar host galaxy. We cannot test this 
possibility further because we do not have narrow-band images in 
filters tuned to $z_{em}=2.543$. 
However, we cannot rule out this possibility either. If O1 is in fact 
the host galaxy of the quasar, then it would have a luminosity (at rest 
frame 0.45 $\mu$m) of $\approx 2.9 \times 10^{10}$ $h_{70}^{-2}$ 
L$_{\odot}$ for $q_{0} =0.5$ or $\approx 6.4 \times 10^{10}$ 
$h_{70}^{-2}$ L$_{\odot}$ for $q_{0} =0.1$.
The images would then suggest that the quasar host is not a 
large galaxy with or without interactions, but rather shows a compact 
morphology. The strongest feature in the quasar host would then be 
off-center with respect to the quasar nucleus, which has been observed 
in other quasars. If O1 is 
in fact the quasar host, then the limits on the luminosity and SFR in 
the DLA are even more severe than our estimates in sections 5.1 and 
5.2. Conversely, if O1 is the DLA galaxy, then the constraints on the 
quasar host are more severe than those given above. 

It is also possible that O1 is an interloper galaxy at an even lower 
redshift than the DLA. However, there is no spectroscopic evidence 
available for this based on the available spectra. Ultraviolet archival 
spectra with HST or IUE (which would contain any potential DLA line at a 
lower redshift) are not available, while the ground-based optical spectra 
are only medium-resolution. We therefore do not consider this possibility further. 

\section{CONCLUSIONS AND FUTURE WORK}

With deep diffraction-limited NICMOS images of LBQS 1210+1731, we have 
probed regions far closer to the quasar sight-line than in 
most previous studies of high-redshift intervening DLAs. The two objects we 
report mark the closest detected high-redshift DLA candidates yet to any 
quasar sight line. 
Our continuum and H$\alpha$ images of the $z=1.89$ DLA toward 
LBQS 1210+1731 suggest that this DLA is not a big 
galaxy with high SFR, but may be compact (2-3 $h_{70}^{-1}$ kpc in 
size), probably consisting of multiple sub-units. Assuming no dust 
extinction of H-$\alpha$ emission, we place a 3 $\sigma$  upper limit of 
4.0 $h_{70}^{-2}$ M$_{\odot}$ yr$^{-1}$ on the star formation rate, 
for $q_{0}=0.5$ . Our continuum and H$\alpha$ observations 
are consistent with the hierarchical models, in which DLAs arise in 
several sub-galactic clumps or dwarf galaxies, which eventually come 
together to form the present-day galaxies (see, e.g., York et al. 1986;  
Matteucci et al. 1997). Indeed, theoretical simulations of 
merging proto-galactic fragments in cold dark matter cosmologies  
(e.g., Haehnelt et al. 1998), low surface brightness galaxies 
(e.g., Jimenez et al. 1999), and collapsing halos with merging clouds  
(e.g., McDonald \& Miralda-Escud'e 1999) have also been found to reproduce 
the observed properties of DLAs (asymmetric line profiles of metal 
absorption lines, metallicities, H I content etc.) 
The small sizes of high-$z$ DLAs suggested by our observations 
are also consistent with the small sizes of galaxies seen in other 
independent high-redshift observations, e.g., in the NICMOS Hubble 
Deep Field observations (Thompson et al. 1999). Together, these 
observations may be indications that while star 
formation had begun long before $z=2$ resulting in some chemical 
enrichment, most of the dynamical assembly of galaxies as we know them 
today occurred more recently, and at $z \sim 2$, the various constituent 
units were still coming together. However, it cannot be ruled out that 
the DLA toward LBQS1210+1731 is a large low surface brightness galaxy 
with a low SFR, which is below our detection limit even in the F160W 
image. 

We point out that our conclusions are, nevertheless, based on detailed 
observations of only one high-$z$ DLA. It is quite possible that 
different DLAs have different rates of evolution because of different 
physical conditions. Indeed, this is suggested by the large scatter in 
the metallicity-redshift relation of 
DLAs (see, e.g., Pettini et al. 1999 and references therein). The 
NICMOS observations of other DLAs from our sample are currently being 
analyzed and will help to explore the generality of our conclusions. 
To improve the statistics of the DLA imaging studies, it is necessary 
to obtain high spatial resolution near-IR images of 
more high-redshift DLAs. It would be very valuable to carry out 
a deeper near-IR imaging survey of more DLAs with HST, if the NICMOS 
cryocooler or the near-IR channel of WFC3 becomes available in 
the near future. A major advantage of such HST observations will be 
a relatively stable PSF compared to that currently achieved with any ground-based telescope, which is crucial for the detection of DLAs. 
It will also be of great interest to complement the HST observations 
with observations from adaptive optics systems on large ground-based 
telescopes. Although these systems will not initially have the 
relatively stable PSF offered by HST, they will be able to achieve 
even higher spatial resolution and higher imaging sensitivity. Such 
future space and ground-based observations will provide further 
insight into the structure and nature of DLA galaxies, and thereby 
help to constrain theoretical models of the formation and evolution of 
galaxies.  

\acknowledgments

This project was supported by NASA grant NAG 5-3042 to the NICMOS 
Instrument Definition Team. It is a pleasure to thank Nicholas Bernstein 
and Keith Noll for their assistance in the scheduling of 
our observations. We thank Elizabeth Stobie, Dyer Lytle, Earl O'Neil, 
Irene Barg, and Anthony Ferro for software and computer support. We also 
thank Andrew Bunker for making his model star formation rate versus 
redshift curves available to us ahead of publication.

\clearpage

\clearpage

\centerline{\bf FIGURE CAPTIONS}

{\bf FIG. 1--} NICMOS camera 2 non-coronagraphic 1.6 $\mu$m broad-band 
image of the field of LBQS 1210+1731. The color scheme is indicated 
with the bar on the bottom of the image. The flux scale in ADU s$^{-1}$ 
is indicated on the color bar. Image Y axis is -121.961 degrees east of 
north.

{\bf FIG. 2--} NICMOS camera 2 non-coronagraphic 
1.9 $\mu$m narrow-band image of the field of 
LBQS 1210+1731. Image Y axis is -121.961 degrees east 
of north.

{\bf FIG. 3--} NICMOS camera 2 coronagraphic 1.6 $\mu$m broad-band 
image of the field of LBQS 1210+1731. 
The quasar has been placed in the coronagraphic hole.
Image Y axis is -123.987 degrees east of north.

{\bf FIG. 4--} Zoomed-in $2.74 $\arcsec$  \times 2.71 $\arcsec$ $ region 
of the NICMOS camera 2 non-coronagraphic 1.6 $\mu$m broad-band 
image of the field of LBQS 1210+1731, (a) before PSF subtraction (top), 
(b) after PSF subtraction (bottom). The residual feature is labeled as O1 
in the bottom panel. 

{\bf FIG. 5--} Zoomed-in $ 2.69 $\arcsec$  \times 2.66 $\arcsec$ $ region 
of the NICMOS camera 2 non-coronagraphic 1.9 $\mu$m narrow-band image of 
the field of LBQS 1210+1731, (a) before PSF subtraction (top), (b) after 
PSF subtraction (bottom). The residual feature is labeled as O1 
in the bottom panel. 

{\bf FIG. 6--} Zoomed-in $2.66 $\arcsec$  \times 2.71 $\arcsec$ $ region 
of the NICMOS camera 2 coronagraphic 1.6 $\mu$m broad-band image of the 
field of LBQS 1210+1731, (a) before PSF subtraction (top), (b) after 
PSF subtraction (bottom). The four residual features are labeled as O2 
in the bottom panel.

{\bf FIG. 7--} Effect of HST ``breathing'' focus variations on the 
PSF subtracted F160W non-coronagraphic image. Zoomed-in 
$2.74 $\arcsec$  \times 2.71 $\arcsec$ $ region of the NICMOS 
camera 2 non-coronagraphic 1.6 $\mu$m broad-band image of the field 
of LBQS 1210+1731, on using images from 4 different observations of 
the PSF star P330E. The PSF star observation dates and breathing values 
are (July 8, 1998; 1.0),  (August 9, 1998; 1.2), 
(September 7, 1998; 0.7), and (March 7, 1998; -1.7) 
respectively for the (a) top left, (b) top right, (c) bottom left, 
and (d) bottom right panels. The quasar observations were obtained on 
July 22, 1998 at breathing value of 2.2.

{\bf FIG. 8--} Differences of PSFs used in Fig. 7. 
(a) PSF for Fig. 7a - PSF for Fig. 7b (top left panel), 
(b) PSF for Fig. 7a - PSF for Fig. 7c (top right), 
(c) PSF for Fig. 7a - PSF for Fig. 7d (bottom left). 

{\bf FIG. 9--} Effect of HST breathing focus variations on the PSF 
subtracted F190N non-coronagraphic image. Zoomed-in 
$3.04 $\arcsec$  \times 3.01 $\arcsec$ $ 
region of the NICMOS camera 2 non-coronagraphic 1.9 $\mu$m narrow-band 
image of the field of LBQS 1210+1731, on using images 
from 3 different observations of the PSF star P330E. The PSF star 
observation dates and breathing values are (May 29, 1998; 1.5), 
(March 7, 1998; 1.3), and (July 8, 1998; 0.7),  
respectively for the (a) top left, (b) top right, and (c) bottom left  
panels. The quasar observations were obtained on July 22, 1998 at mean 
breathing value of 2.3.

{\bf FIG. 10--} Differences of PSFs used in Fig. 9. (a) PSF for Fig. 
9a - PSF for Fig. 9b (top left panel), (b) PSF for Fig. 9a - PSF for 
Fig. 9c (top right). 

{\bf FIG. 11--} Effect of using different PSF stars on the PSF 
subtracted F160W non-coronagraphic image. Zoomed-in 
$2.74 $\arcsec$  \times 2.71 $\arcsec$ $ 
region of the NICMOS camera 2 non-coronagraphic 1.6 $\mu$m broad-band 
image of the field of LBQS 1210+1731, on using PSF stars P330E, 
BRI 0021, Q1718PSF, and GSC4. The PSF star 
observation dates and average breathing values are (July 8, 1998; 1.0), 
(December 20, 1997; -1.2), (July 21, 1998; 1.8) and 
(Nov. 11, 1997; -3.2),  
respectively for the (a) top left, (b) top right, (c) bottom left, 
and (d) bottom right   
panels. The quasar observations were obtained on July 22, 1998 at mean 
breathing value of 2.3. The differences between the different panels 
may be largely because of breathing differences. (The match with 
the quasar breathing value is poor for panels b and d, while 
it is best for panel c.) The negative feature  near the right edge of 
panel (c) is because of a second star near 
the main PSF star used for subtraction. 

{\bf FIG. 12--} Differences of PSFs used in Fig. 11. (a) PSF for Fig. 
11a - PSF for Fig. 11b (top left panel), (b) PSF for Fig. 11a - PSF for 
Fig. 11c (top right), and (c) PSF for Fig. 11a - PSF for Fig. 11d 
(bottom left). The negative feature near the right edge of
 panel (b) is because of a second star near 
the main PSF star used for subtraction in panel (c) of Fig. 11. 
 
{\bf FIG. 13--} Effect of using different individual dither positions 
on the PSF subtracted F160W non-coronagraphic image. Zoomed-in 
$3.12 $\arcsec$  \times 3.09 $\arcsec$ $ 
region of the field of LBQS 1210+1731, on using position 1, position 4, 
position 5 in the spiral-dither pattern (top left, top right and 
bottom left panels respectively). The bottom right panel shows 
the result of combining the three positions. 

{\bf FIG. 14--} Effect of different focus positions 
on the PSF subtracted F160W non-coronagraphic image. Zoomed-in 
$3.12 $\arcsec$  \times 3.09 $\arcsec$ $ 
region of the NICMOS camera 2 non-coronagraphic 1.6 $\mu$m broad-band 
image of the field of LBQS 1210+1731. (a) The top left 
panel shows the PSF subtraction obtained on using PSF star 
image from the field of galaxy cluster CL0939+47 which has the same 
``camera 1-2'' focus as our quasar data. (b) The top right panel shows  
the difference of two simulated Tiny Tim 
PSFs corresponding to the ``camera 2 focus'' and ``camera 1-2'' focus. 
(c) The bottom left panel shows, on the same intensity stretch as 
the top left panel, the difference of the two simulated Tiny Tim PSFs 
after normalizing each to match the quasar.
(d) The bottom right panel shows the quasar image after subtracting 
a synthetic PSF made by multiplying the July 8, 1998 image of P330E  
by the ratio of the Tiny Tim models for the two focii. See the text 
for details. 

{\bf FIG. 15--} Effect of different magnification 
schemes in data reduction and different interpolation schemes in image 
display on the PSF subtracted F160W non-coronagraphic 
image. Zoomed-in $3.12 $\arcsec$  \times 3.09 $\arcsec$ $ 
region of the F160W image for (a) no magnification in image 
analysis, bicubic interpolation in image display (top left), 
(b) magnification by a factor of 2 and bicubic interpolation in image 
display (top right), (c) no magnification and pixel replication in 
image display (bottom left), (d) magnification by a factor of 2 and 
pixel replication in image display (bottom right). Note the 
similarities between the magnified and unmagnified 
images, resulting from camera 2 being almost critically 
sampled at 1.6 $\mu$m.

{\bf FIG. 16--} Effect of different magnification 
schemes in data reduction and different interpolation schemes in image 
display on the PSF subtracted F190N non-coronagraphic 
image. Zoomed-in $3.12 $\arcsec$  \times 3.09 $\arcsec$ $ 
region of the F190N image for (a) no magnification in image 
analysis, bicubic interpolation in image display (top left), 
(b) magnification by a factor of 2 and bicubic interpolation in image 
display (top right), (c) no magnification and pixel replication in 
image display (bottom left), (d) magnification by a factor of 2 and 
pixel replication in image display (bottom right). 

{\bf FIG. 17--} Effect of including or excluding the image core  
on the PSF subtracted F160W non-coronagraphic image. Zoomed-in 
$2.74 $\arcsec$  \times 2.71 $\arcsec$ $ region of the F160W  
image of the field of LBQS 1210+1731 obtained on (a) including 
the region indicated by the circular mask (left panel) and (b) 
excluding the region indicated by the circular mask (right panel). 
Note the similarity between the two panels.

{\bf FIG. 18--} Effect of shifting the PSF star by 0.1 pixel 
in various directions relative to the quasar on the 
PSF subtracted non-coronagraphic F160W image. The central 
panel (e) is the optimum minimum-variance solution (same as 
Fig. 4b). Top central (b) and bottom central (h) panels correspond 
to PSF star shifts of 0.1 pixel in +y and -y directions. Left central 
(d) and right central (f) panels correspond to PSF star shifts 
of 0.1 pixel in -x and +x directions. Top left (a), top right (c), 
bottom left (g), and bottom right (i) panels correspond to PSF star 
shifts of 0.1 pixel in the top left, top right, bottom 
left, and bottom right directions, respectively. The 
large residuals caused by the slight shifts illustrate 
how well centered the PSF star is with respect to the quasar in the 
optimum PSF subtraction. The regions shown are 2.74 
$\arcsec \, \times 2.71 \arcsec$ regions around the quasar. 

{\bf FIG. 19--} Mean star formation rate in DLAs in 
M$_{\odot}$ yr$^{-1}$ as a function of redshift, for $q_{0}= 0.5$, 
$H_{0} = 70$ km s$^{-1}$ Mpc$^{-1}$. The filled 
triangle shows the upper limit from this 
work, while the unfilled triangles show the limits from Bunker 
et al. (1999). The curve shows the prediction from a closed-box model 
applied to proto-disk galaxies, as calculated by Bunker et al. 
(1999). Note that our SFR limit is a factor of 
3 improvement over the tightest limits of Bunker et al. (1999), and 
that most data points are inconsistent with the proto-disk model. 

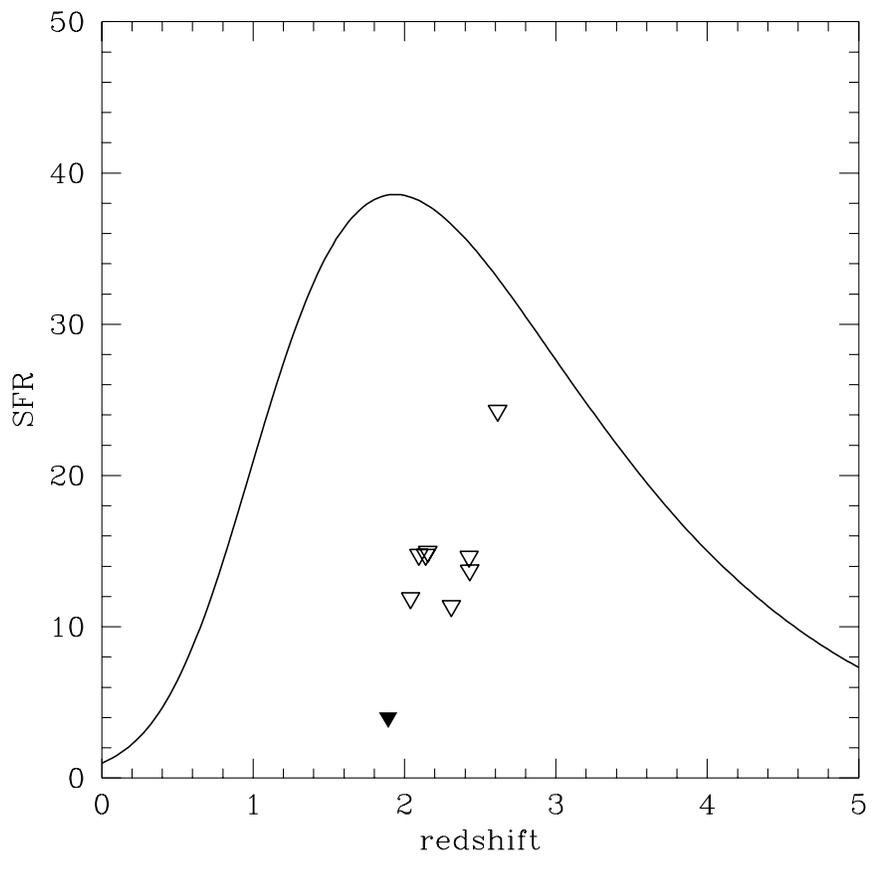
\end{document}